\documentclass{SciPost}

\hypersetup{
    colorlinks,
    linkcolor={red!50!black},
    citecolor={blue!50!black},
    urlcolor={blue!80!black}
}

\usepackage{subcaption}
\usepackage{pdfpages}
\usepackage{hyperref}
\usepackage[bitstream-charter]{mathdesign}
\DeclareSymbolFont{usualmathcal}{OMS}{cmsy}{m}{n}
\DeclareSymbolFontAlphabet{\mathcal}{usualmathcal}

\fancypagestyle{SPstyle}{
\fancyhf{}
\lhead{\colorbox{scipostblue}{\bf \color{white} ~SciPost Physics Proceedings }}
\rhead{{\bf \color{scipostdeepblue} ~Submission }}

\fancyfoot[C]{\textbf{\thepage}}
}

\begin{document}
\renewcommand\thefootnote{}
\footnotetext{\textcopyright~Copyright 2024 CERN for the benefit of the ATLAS Collaboration. Reproduction of this article or parts of it is allowed as specified in the CC-BY-4.0 license.}
\renewcommand\thefootnote{\arabic{footnote}}

\pagestyle{SPstyle}

\begin{center}{\Large \textbf{\color{scipostdeepblue}{
ATLAS EFT Results in the Top Quark Sector\\
}}}\end{center}

\begin{center}\textbf{
Dongwon Kim\textsuperscript{1,2} on behalf of the ATLAS Collaboration
}\end{center}

\begin{center}
{\bf 1} Stockholm University, Sweden
\\
{\bf 2} Oskar Klein Centre, Sweden
\\[\baselineskip]
E-mail: \href{mailto:email1}{\small dong.won.kim@cern.ch}\,
\end{center}

\definecolor{palegray}{gray}{0.95}
\begin{center}
\colorbox{palegray}{
  \begin{tabular}{rr}
  \begin{minipage}{0.36\textwidth}
    \includegraphics[width=60mm,height=1.5cm]{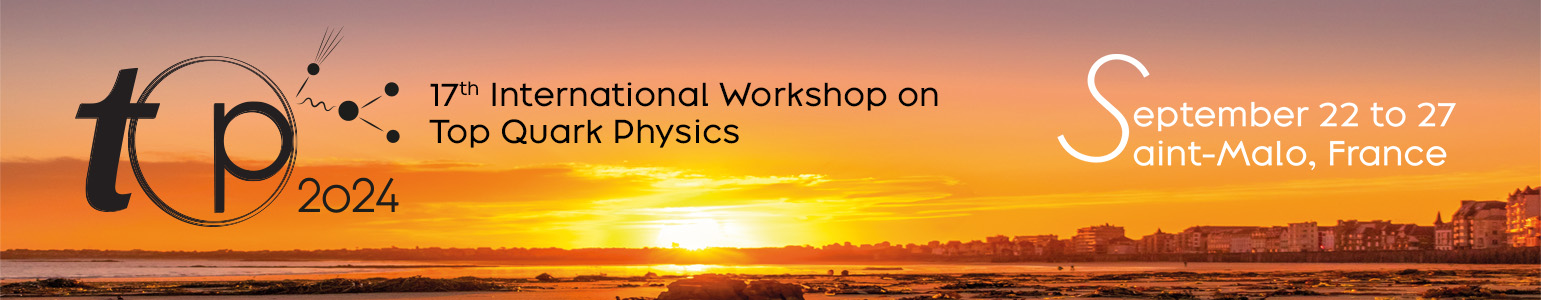}
  \end{minipage}
  &
  \begin{minipage}{0.55\textwidth}
    \begin{center} \hspace{5pt}
    {\it The 17th International Workshop on\\ Top Quark Physics (TOP2024)} \\
    {\it Saint-Malo, France, 22-27 September 2024
    }
    \doi{10.21468/SciPostPhysProc.?}\\
    \end{center}
  \end{minipage}
\end{tabular}
}
\end{center}

\section*{\color{scipostdeepblue}{Abstract}}
\textbf{\boldmath{%
With no evidence of direct production of beyond the Standard Model (BSM) particles at the TeV scale, deviations from the Standard Model (SM) can be explored systematically through Effective Field Theories (EFTs) such as the Standard Model EFT (SMEFT). The SMEFT, a framework for probing BSM effects, extends the SM by introducing higher-dimensional operators parameterized by Wilson coefficients. This contribution highlights three recent analyses using the ATLAS Run-2 dataset at a center-of-mass energy of $\sqrt{s}$ = 13 TeV with an integrated luminosity of 140 $\text{fb}^{-1}$. Combined measurements enhance sensitivity to Wilson coefficients, exploring the potential of SMEFT in the top quark sector.
}}

\vspace{\baselineskip}

\noindent\textcolor{white!90!black}{%
\fbox{\parbox{0.975\linewidth}{%
\textcolor{white!40!black}{\begin{tabular}{lr}%
  \begin{minipage}{0.6\textwidth}%
    {\small Copyright attribution to authors. \newline
    This work is a submission to SciPost Phys. Proc. \newline
    License information to appear upon publication. \newline
    Publication information to appear upon publication.}
  \end{minipage} & \begin{minipage}{0.4\textwidth}
    {\small Received Date \newline Accepted Date \newline Published Date}%
  \end{minipage}
\end{tabular}}
}}
}




\section{Introduction}
\label{sec:intro}
No evidence of direct production of BSM particles at the TeV scale by the ATLAS experiment \cite{ATLAS:2008xda} suggests that effects from physics beyond the Standard Model (BSM) could arise at energy scales beyond the current experimental reach. Small deviations from Standard Model (SM) predictions, however, can still be systematically explored using Effective Field Theories (EFTs), such as the Standard Model EFT (SMEFT). SMEFT extends the SM by introducing higher-dimensional operators parametrized by Wilson coefficients, which provide a framework for quantifying potential BSM effects and precisely measuring quantum interference between standard and effective interactions \cite{Brivio:2017vri,Englert:2014cva,Buchmuller:1985jz} in the following way:
\begin{equation*}
\mathcal{L}_{\text{SMEFT}} = \mathcal{L}_{\text{SM}} + \sum_{i, d>4} \frac{c_i^{(d)} \mathcal{O}_i^{(d)}}{\Lambda^{d-4}} = \mathcal{L}_{\text{SM}} + \sum_{i} \frac{c_i^{(6)} \mathcal{O}_i^{(6)}}{\Lambda^{2}} + \sum_{j} \frac{c_j^{(8)} \mathcal{O}_j^{(8)}}{\Lambda^{4}} + \cdots
\end{equation*}
where $\mathcal{L}_{\text{SMEFT}}$, $\mathcal{L}_{\text{SM}}$, $c_i$, $d$, $\mathcal{O}_i$, and $\Lambda$ represent the SMEFT Lagrangian, the SM Lagrangian, the Wilson coefficients, the operator dimension, the higher-dimensional operators, and the cut-off scale, respectively.  As the heaviest Standard Model particle, the top quark is particularly sensitive to BSM physics, rare processes involving top quarks are essential probes in SMEFT studies. In this context, EFT interpretations of three recent ATLAS results based on the full Run-2 dataset acquired during 2016-2018 and corresponding to an integrated luminosity of 140 $\text{fb}^{-1}$ are highlighted: measurements of $t\bar{t}$ production in association with a photon ($\gamma$) \cite{ATLAS:2024hmk}, $tq$ production in the t-channel \cite{ATLAS:2024ojr}, and searches for charged lepton-flavour-violating (cLFV) interactions involving $\mu \tau q t$ \cite{ATLAS:2024njy}, where $q$ denotes either $u$- or $c$-quark.

\section{Top quark measurements and searches with EFT interpretation}
\label{sec:EFTresults}
Measurements of different processes provide sensitivity to distinct EFT operators. In ATLAS EFT, analyses of top-quark processes primarily consider dimension-six operators ($\mathcal{O}^{(6)}$), but results are often compared against scenarios that include both SM-EFT interference ($\Lambda^{-2}$) and higher-order EFT-EFT interference ($\Lambda^{-4}$) terms to account for the potential impact of higher-dimensional contributions. The SMEFT operators are defined using the Warsaw basis and implemented through Monte Carlo tools like \textit{MadGraph5\_aMC@NLO}\cite{Alwall:2014hca} and dedicated packages such as \textit{SMEFTatNLO}\cite{Degrande:2020evl}, \textit{dim6top}\cite{Aguilar-Saavedra:2018ksv}, and \textit{TopFCNC}\cite{Degrande:2014tta}. For these studies, the new physics scale is typically fixed at $\Lambda = 1 \ \text{TeV}$, and constraints are placed on both individual and marginalised Wilson coefficients. Generally, only a small set of operators impacting relevant vertices is analyzed, though some measurements also provide joint constraints on pairs of coefficients.
\subsection{Measurement of inclusive and differential $t\bar{t}\gamma$ cross sections}
Inclusive and differential cross-section measurements of $t\bar{t}\gamma$ production with SMEFT interpretations are presented, utilizing single-lepton and dilepton final states. The fiducial phase space considers photons radiated either from an initial-state parton or from one of the top quarks. For the single-lepton channel, a multi-class neural network (NN) is employed to separate signal from background, while a binary classifier is used in the dilepton channel. The results demonstrate good agreement with the SM, and it was carefully checked that the input variables are well-modelled in the data.

\begin{figure}
    \centering
    \includegraphics[width=0.5\linewidth]{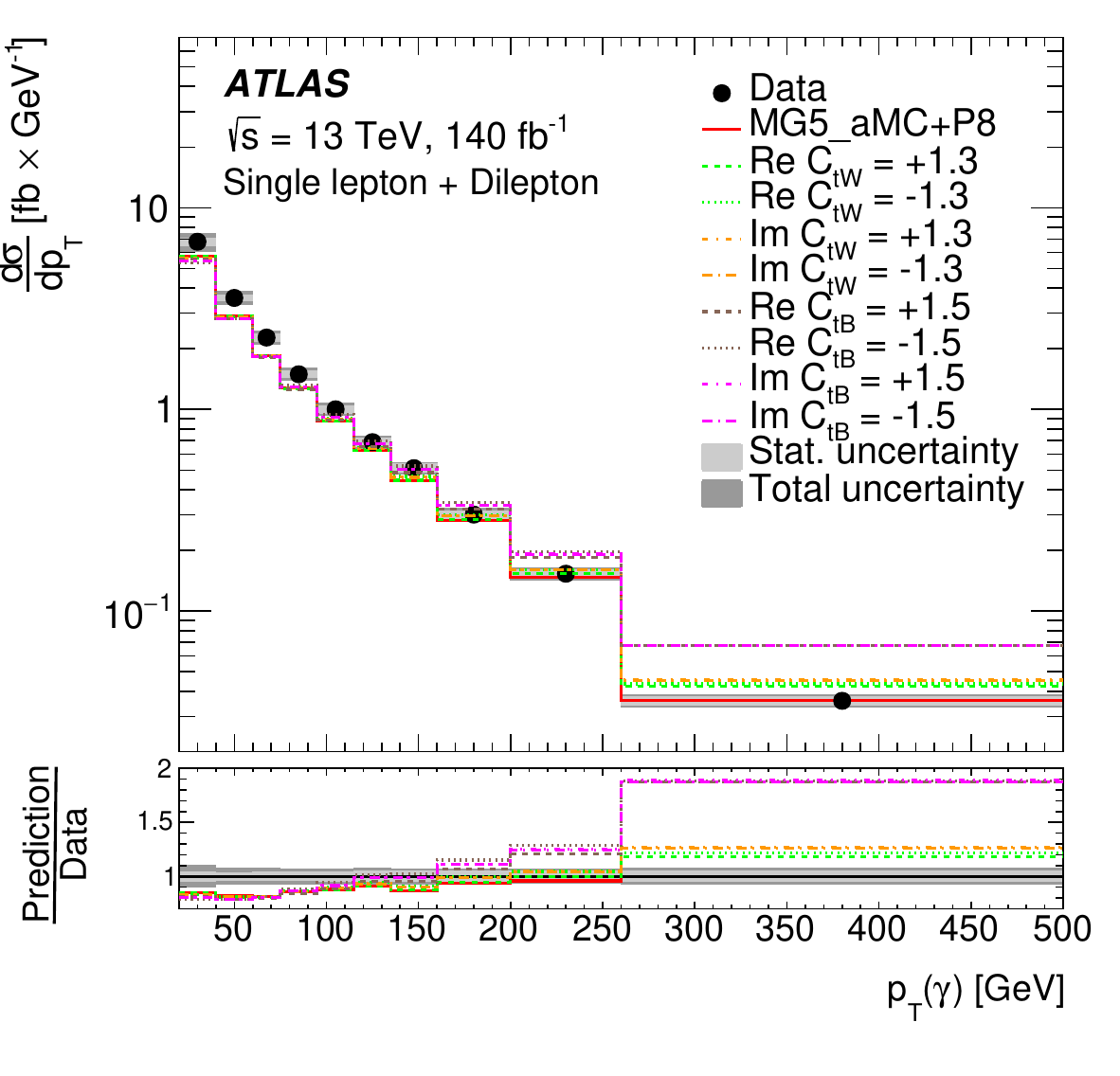}
    \caption{Comparison of the photon $p_\mathrm{T}$ distribution from the combined measurement in the single-lepton and dilepton channels with the SM prediction and scenarios with one non-zero EFT parameter. The lower panel shows the ratio of predictions to the data. \cite{ATLAS:2024hmk}}
    \label{fig:ttgamma_photon_pt_distribution}
\end{figure}

\begin{figure}[h!]
    \centering
    \begin{subfigure}[t]{0.45\linewidth}
        \centering
        \includegraphics[width=\linewidth]{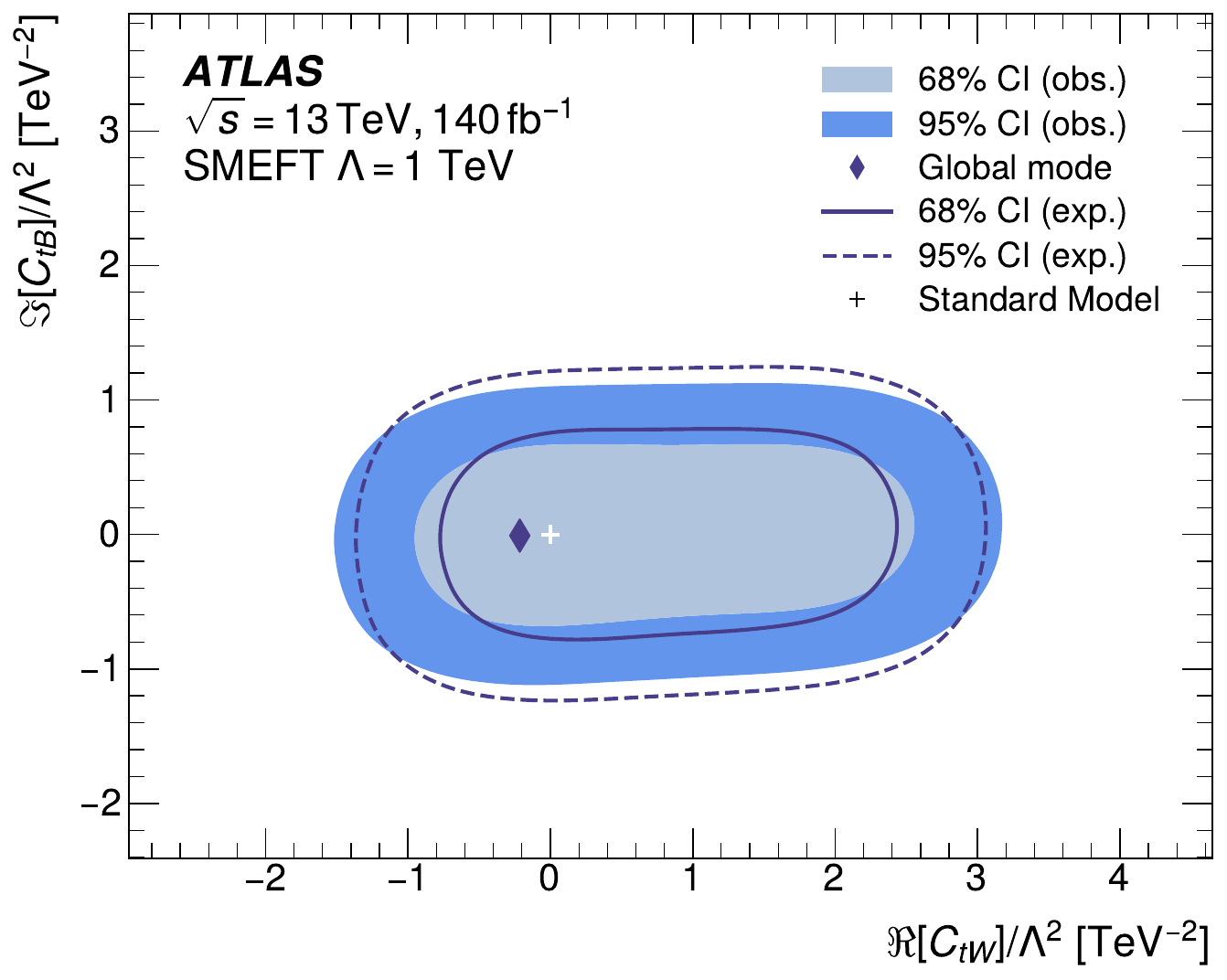}
        \caption{$\mathrm{Re}[C_{tW}]$ vs. $\mathrm{Im}[C_{tB}]$ from $t\bar{t}\gamma$ measurements alone.}
    \end{subfigure}
    \hspace{0.05\linewidth}
    \begin{subfigure}[t]{0.45\linewidth}
        \centering
        \includegraphics[width=\linewidth]{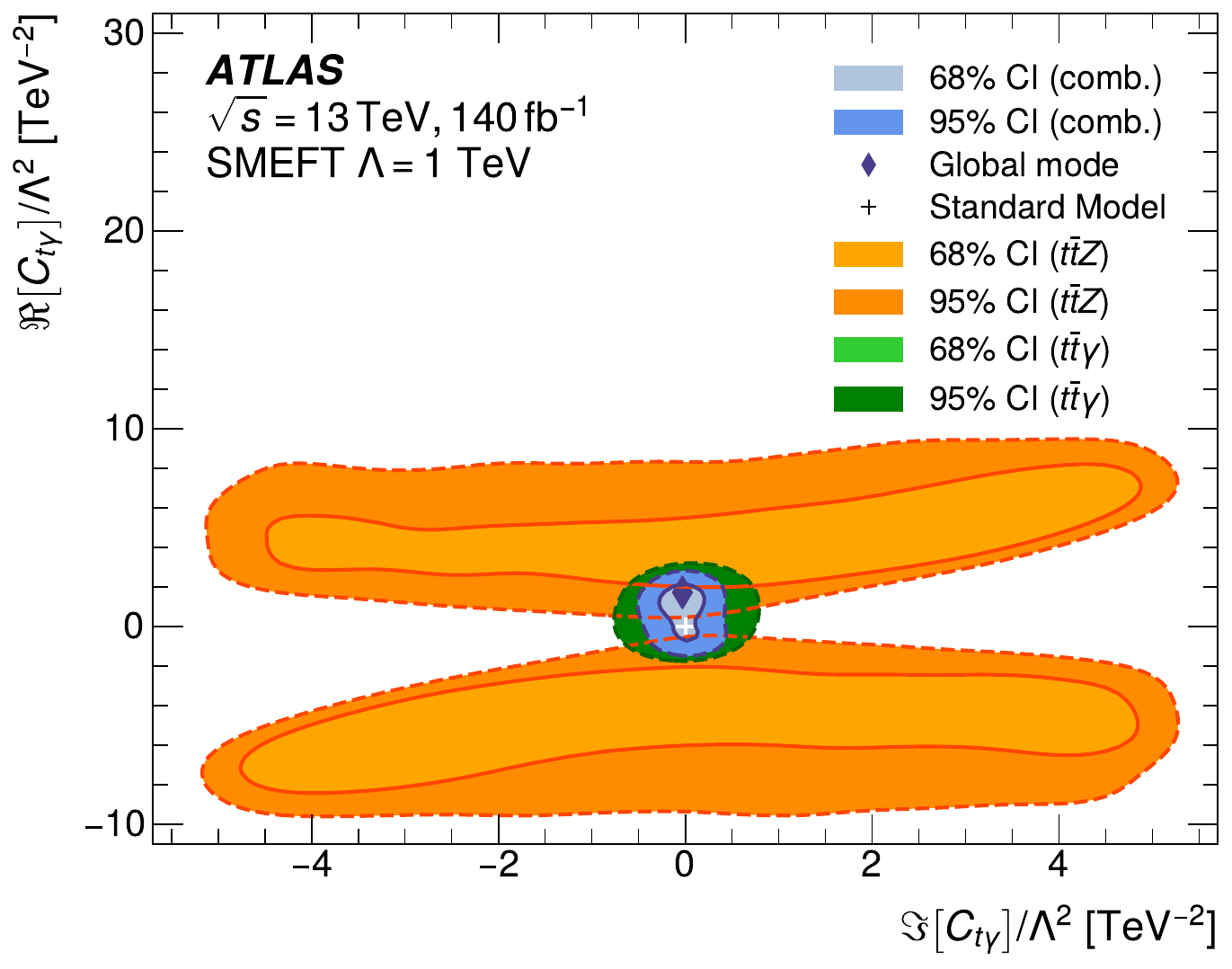}
        \caption{$\mathrm{Im}[C_{t\gamma}]$ vs. $\mathrm{Re}[C_{t\gamma}]$ from $t\bar{t}\gamma$ and $t\bar{t}Z$ measurements.}
    \end{subfigure}
    \caption{Two-dimensional marginalized posteriors for the EFT operators from the quadratic fits, indicating 68\% and 95\% confidence intervals. \cite{ATLAS:2024hmk}}
    \label{fig:eft_ttgamma}
\end{figure}

A SMEFT interpretation is performed using the differential $t\bar{t}\gamma$ measurements, with additional insights gained by combining $t\bar{t}\gamma$ and $t\bar{t}Z$ data. Both the real and imaginary components of the Wilson coefficients $\mathcal{C}_{tB}$ and $\mathcal{C}_{tW}$ are extracted, including linear, cross-term, and quadratic contributions. This is achieved through a profile likelihood fit to the photon transverse momentum ($p_\mathrm{T}$) distribution at the particle level in Figure \ref{fig:ttgamma_photon_pt_distribution}. Extracted limits on $\mathcal{C}_{tB}$ and $\mathcal{C}_{tW}$ are consistent with SM predictions, as shown in Figure \ref{fig:eft_ttgamma}(a).

Since $\mathcal{C}_{tB}$ and $\mathcal{C}_{tW}$ also influence $t\bar{t}Z$ production, the combination of $t\bar{t}\gamma$ and $t\bar{t}Z$ provides stronger constraints, particularly on $\mathcal{C}_{tW}$. Furthermore, combining multiple measurements offers improved discrimination power, addressing degeneracies in limit extractions (e.g., Figure \ref{fig:eft_ttgamma}(b), where $\mathcal{C}_{t\gamma}$ and $\mathcal{C}_{tZ}$ are linear combinations of $\mathcal{C}_{tB}$ and $\mathcal{C}_{tW}$). No significant deviations from the SM are observed,  within the limits at 95\% confidence level (CL).

\subsection{Measurement of $tq$($\bar{t}q$) cross sections in t-channel}
\begin{figure}[h!]
    \centering
    \begin{subfigure}[t]{0.45\linewidth}
        \centering
        \includegraphics[width=\linewidth]{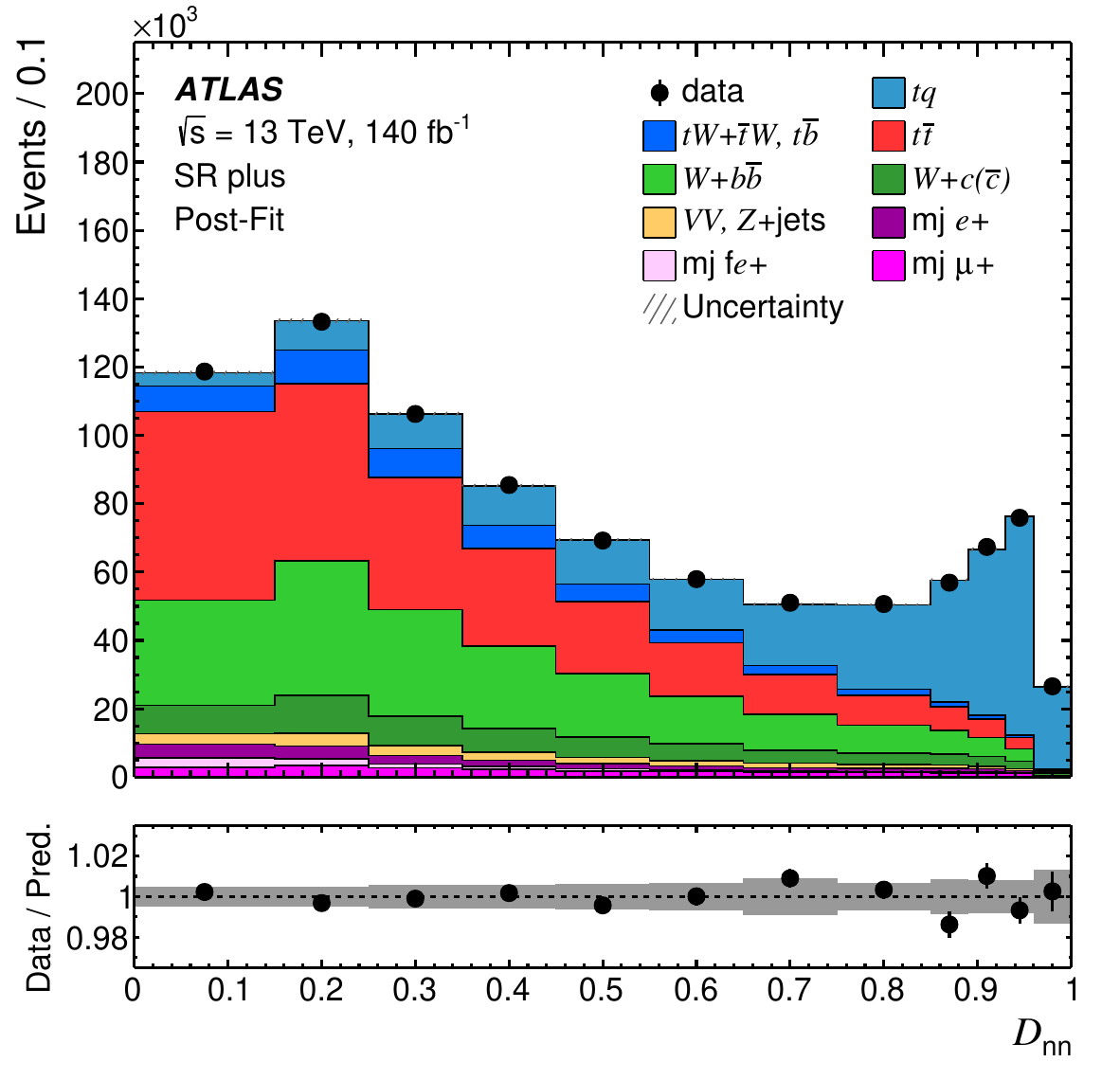}
        \caption{SR plus (with $l^{+}$)}
    \end{subfigure}
    \hspace{0.05\linewidth}
    \begin{subfigure}[t]{0.45\linewidth}
        \centering
        \includegraphics[width=\linewidth]{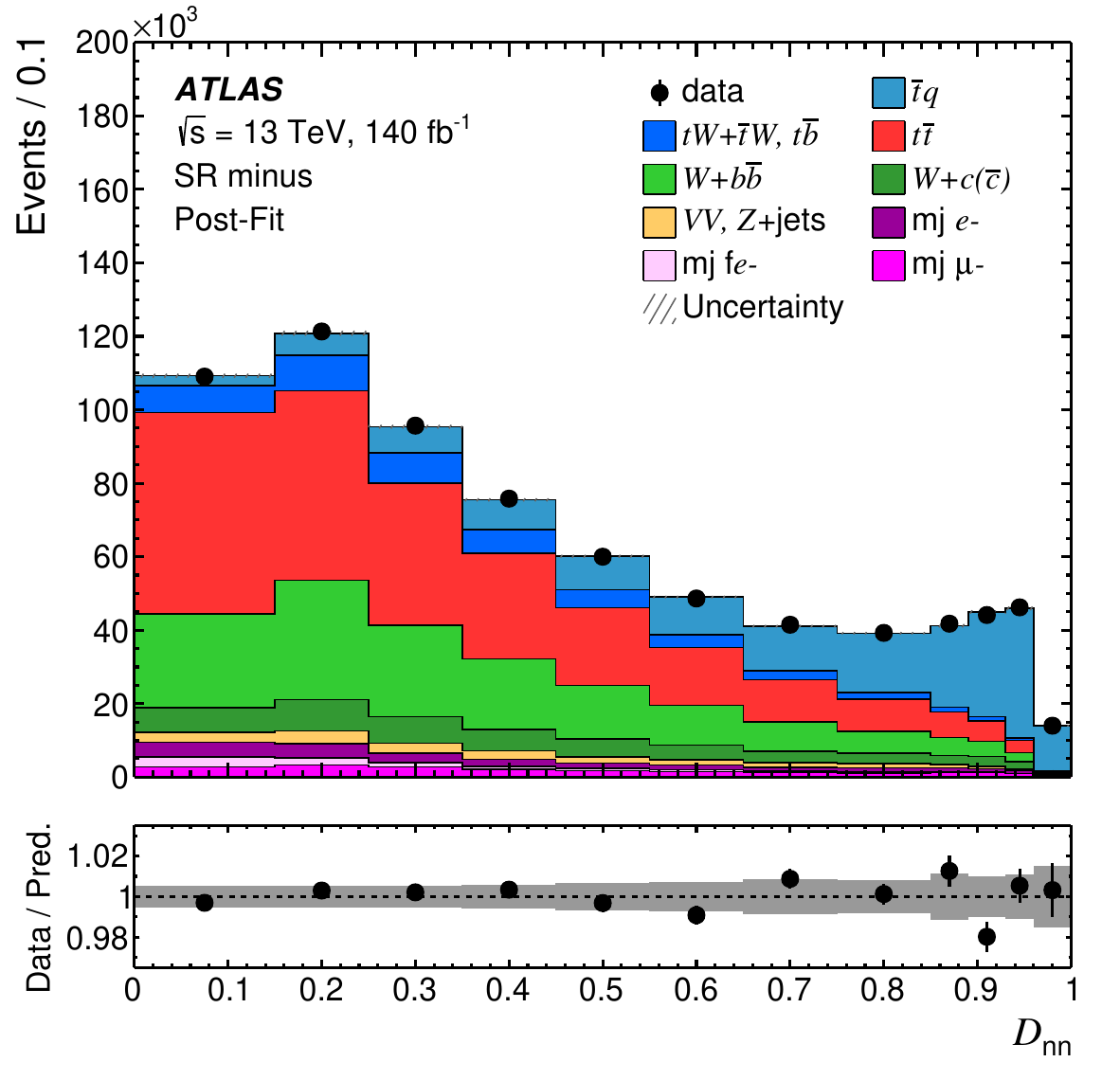}
        \caption{SR minus (with $l^{-}$)}
    \end{subfigure}
    \caption{Observed $D_\mathrm{nn}$ post-fit distributions (black markers) for (a) SR plus and (b) SR minus. The hatched band overlaid on the stacked histograms includes all uncertainties on the SM prediction, accounting for correlations from the fit. The lower panel shows the data-to-prediction ratio with uncertainties as a gray band. \cite{ATLAS:2024ojr}}
    \label{fig:eft_singletop}
\end{figure}

The cross-section measurements of the production of a single top-quark or top-antiquark via the t-channel exchange of a virtual $W$ boson are performed. Events are selected to include either one isolated electron or muon, significant missing transverse energy ($E_\mathrm{T}^\mathrm{miss}$), and exactly two hadronic jets with high $p_\mathrm{T}$, one of which is required to be $b$-tagged. An artificial NN was employed to construct a discriminant that efficiently separates signal from background events. The distributions of the NN discriminant ($D_\mathrm{nn}$) in Figure \ref{fig:eft_singletop} are used in profile maximum-likelihood fits to extract the signal yields. These measurements are interpreted in the framework of SMEFT, setting 95\% CL constraints on the Wilson coefficients of two operators: the four-quark operator, $C_{Qq}^{3,1}/\Lambda^2$ (-0.37 < $C_{Qq}^{3,1}/\Lambda^2$ < 0.06), and the operator coupling the third quark generation to the Higgs boson doublet, $C_{\phi Q}^3/\Lambda^2$ (-0.87 < $C_{\phi Q}^3/\Lambda^2$ < 1.42). Fully simulated samples incorporating SMEFT effects have been used to ensure accurate modelling of these contributions.

\subsection{Search for charged-lepton-flavour violating (cLFV) in $\mu \tau q t$ interactions}
\begin{figure}[h!]
    \centering
    \includegraphics[width=0.5\linewidth]{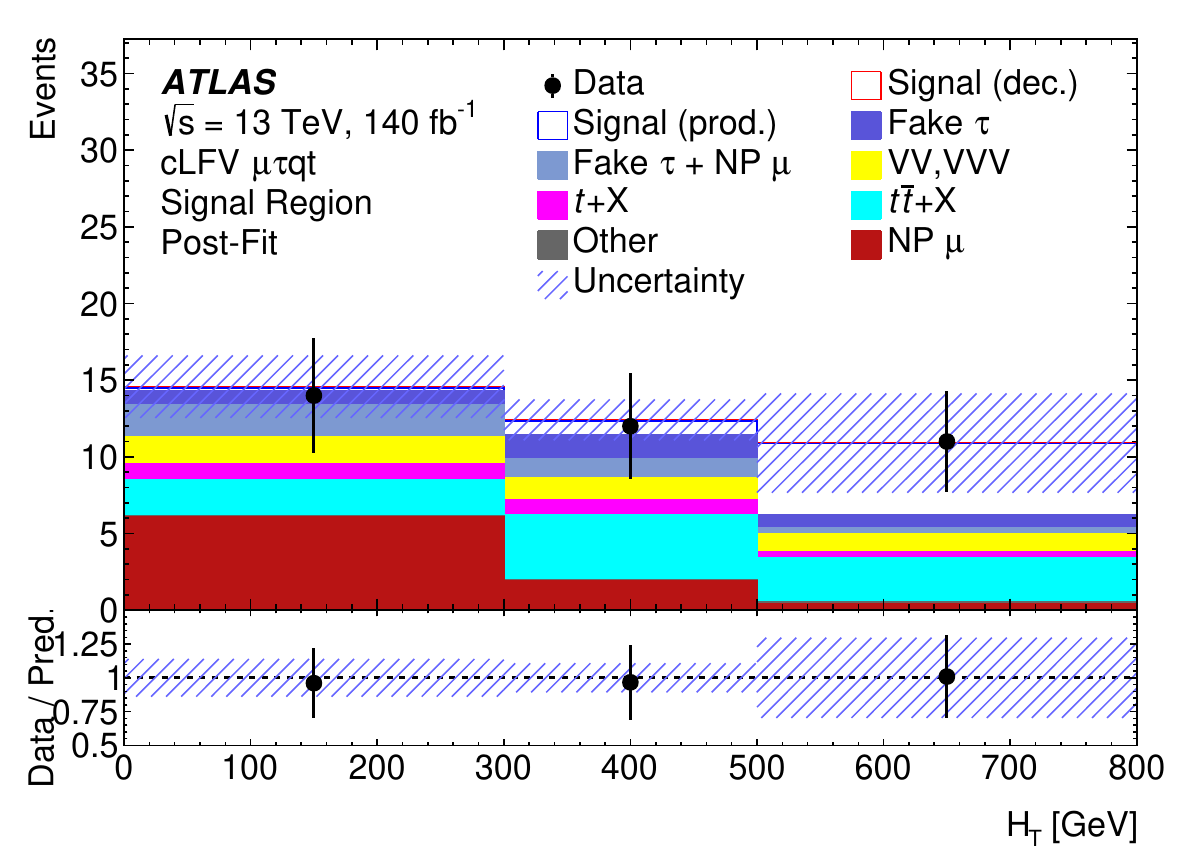}
    \caption{Observed event yields in the SR compared with MC simulations as a function of the scalar sum of the lepton and jet transverse momenta $H_\mathrm{T}$. The signal sample is generated with all Wilson coefficients simultaneously set to 0.1 for a new physics scale of $\Lambda$ = 1 TeV. \cite{ATLAS:2024njy}}
    \label{fig:CLFV_EFT_SR}
\end{figure}

A search for cLFV interactions in  $\mu \tau q t$ processes is conducted, considering both top-quark production and decay. Events with two same-sign leptons, a hadronically decaying $\tau$-lepton, and at least one  $b$-tagged jet are considered (See Figure \ref{fig:CLFV_EFT_SR}). No significant excess over the SM prediction is observed, and 95\% $\text{CL}_S$ limits are set on the branching ratio of $\mathcal{B}(t \to \mu \tau q) < 8.7 \times 10^{-7}$. SMEFT interpretations are performed, yielding 95\% CL constraints on the Wilson coefficients, depending on the flavour of the associated light quark and the Lorentz structure of the coupling. These range from $|C_{\text{lequ}}^{3(2313)}| / \Lambda^2 < 0.10 \, \text{TeV}^{-2}$ for  $\mu \tau u t$ to $|C_{\text{lequ}}^{3(2323)}| / \Lambda^2 < 1.8 \, \text{TeV}^{-2}$ for $\mu \tau c t$ (See Table \ref{tab:CLFV_limits}). These results complement previous EFT analyses \cite{Chala:2018agk} and provide further constraints on cLFV processes involving the top quark.

\begin{table}[h!]
    \centering
    \includegraphics[width=0.65\linewidth]{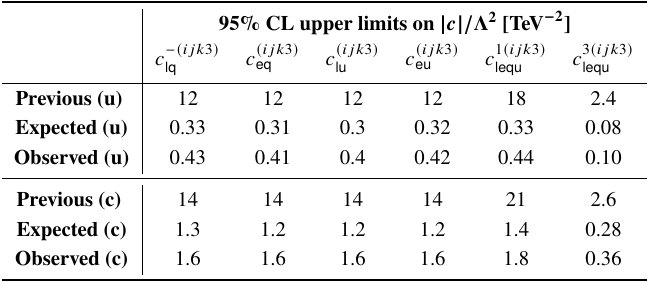}
    \caption{Expected and observed 95\% CL upper limits on Wilson coefficients corresponding to 2Q2L EFT operators that could introduce cLFV top decay in the $\mu \tau$ channel \cite{ATLAS:2024njy} with existing limits from Ref.~\cite{Chala:2018agk}.}
    \label{tab:CLFV_limits}
\end{table}

\section{Conclusion}
Three analyses utilizing the full Run-2 dataset of the ATLAS detector have been summarized, focusing on searches for BSM physics in the top quark sector through an EFT framework. The combination of improved measurements and the inclusion of new channels has significantly enhanced the sensitivity to Wilson coefficients by breaking degeneracies and tightening constraints. Furthermore, since many EFT-based analyses at ATLAS assume $\Lambda$ of 1 TeV, near the current experimental energy scale, the collaboration has begun setting direct limits on $\Lambda$, exploring scenarios with varying coupling strengths ($c_i$ = 0.01, 1, 4$\pi^2$)\cite{LHCTopWGSummaryPlots}. In the top quark sector, SMEFT continues to adapt and advance, aligning with the increasing precision of experimental data from the LHC.

\bibliography{SciPost_Example_BiBTeX_File.bib}

\end{document}